\documentclass[conference]{IEEEtran}
\usepackage[submission]{main_style}
\usepackage{times}
\usepackage{helvet}
\usepackage{courier}
\usepackage[hyphens]{url}
\usepackage{graphicx}
\usepackage{caption}
\usepackage{algorithm}
\usepackage{algorithmic}
\usepackage{tabularx}
\usepackage{multirow}

\usepackage{float}
\usepackage{newfloat}
\usepackage{listings}
\DeclareCaptionStyle{ruled}{labelfont=normalfont,labelsep=colon,strut=off}
\floatstyle{ruled}
\newfloat{listing}{tb}{lst}{}
\floatname{listing}{Listing}

\title{The Self-Perception and Political Biases of ChatGPT}
\author {
    Jérôme Rutinowski,
    Sven Franke,
    Jan Endendyk,
    Ina Dormuth,
    Markus Pauly
}
\affiliations {
    TU Dortmund University,  Joseph-von-Fraunhofer-Straße 2-4, 44227 Dortmund, Germany\\
    Correspondence: jerome.rutinowski@tu-dortmund.de\\
}

\usepackage{bibentry}

\begin{document}

\maketitle

\thispagestyle{plain}
\pagestyle{plain}

\begin{abstract}

This contribution analyzes the self-perception and political biases of OpenAI's Large Language Model ChatGPT.
Taking into account the first small-scale reports and studies that have emerged, claiming that ChatGPT is politically biased towards progressive and libertarian points of view, this contribution aims to provide further clarity on this subject.
For this purpose, ChatGPT was asked to answer the questions posed by the political compass test as well as similar questionnaires that are specific to the respective politics of the G7 member states.
These eight tests were repeated ten times each and revealed that ChatGPT seems to hold a bias towards progressive views.
The political compass test revealed a bias towards progressive and libertarian views, with the average coordinates on the political compass being \mbox{(-6.48, -5.99)} (with (0, 0) the center of the compass, i.e., centrism and the axes ranging from -10 to 10), supporting the claims of prior research.
The political questionnaires for the G7 member states indicated a bias towards progressive views but no significant bias between authoritarian and libertarian views, contradicting the findings of prior reports, with the average coordinates being \mbox{(-3.27, 0.58)}.
In addition, ChatGPT's Big Five personality traits were tested using the OCEAN test and its personality type was queried using the Myers–Briggs Type Indicator (MBTI) test.
Finally, the maliciousness of ChatGPT was evaluated using the Dark Factor test.
These three tests were also repeated ten times each, revealing that ChatGPT perceives itself as highly open and agreeable, has the Myers-Briggs personality type ENFJ, and is among the 15\% of test-takers with the least pronounced dark traits.

\end{abstract}

\section{Introduction} \label{sec:1}

Recently, Large Language Models (LLMs) have gained tremendous amounts of attention by experts as well as the general public.
A notable example of one such model is \mbox{OpenAI's} ChatGPT (Generative Pre-trained Transformer).
ChatGPT is a model that generates text responses, when a user provides it with a prompt.
It is an LLM that was fine-tuned based upon sophisticated machine learning techniques and human feedback.
Currently, ChatGPT is open-access (version 3.5 is free to use, while version 4 is available as a subscription service) but not open-source. 
Due to this, users can only make assumptions as to why it behaves the way it does and what data it might have been trained on, with the developers claiming that it was trained on “[...] vast amounts of data from the internet written by humans, including conversations~[...]"~\cite{openai}.
While it receives a lot of positive acclaim and often seems to work as intended, prominent figures such as Yann LeCun and Yoshua Bengio have criticized it publicly for various reasons, one of them being, that LLMs might not be the right approach towards AGI (Artificial General Intelligence) \cite{lecun23,openletter2023}.
A reason for which users have been criticizing the model is for its supposed bias towards progressive and libertarian views, claiming that an AI model should not hold such biases \cite{mcgee2023chat}.

In this work we investigate these claims and study ChatGPT's political biases. 
We additionally investigate whether ChatGPT's self-perception is such that it can be attributed personality traits based on commonly used psychological assessments.
We subsequently investigate whether there is a relationship between personality traits and ChatGPT's political biases.
In the following section, we discuss the relevant literature related to this contribution.
Subsequently, we present our methodology and then finally analyze the results of our experiments and draw a conclusion based on them.

\section{Related Work} \label{sec:2}

This section provides the reader with a brief insight into the workings of ChatGPT's functioning.
It also presents a set of measures for political biases and personality traits and how these were already applied to ChatGPT in previous publications.

\subsection{Large Language Models} \label{sec:2.1}

The term “Large Language Model" is an umbrella term for language (generation) neural network architectures that are trained on large amounts of unlabeled data, e.g., as self-supervised Pretrained Foundation Models (PFMs) \cite{manning2022human,zhou2023comprehensive}.
For OpenAI's ChatGPT, for instance, this results in a model with a total of 1.5~billion hyperparameters for ChatGPT-2, to 175 billion hyperparameters for ChatGPT-3, and a currently undisclosed amount of hyperparameters for ChatGPT-4.
What is known is that ChatGPT uses a transformer architecture, which is an architecture that was developed as an alternative to recurrent neural networks by Google and the University of Toronto \cite{vaswani2017attention}.
Transformers use a typical encoder and decoder architecture that can parse sequence data.
The key features of transformers are their positional encoding and self-attention functionalities, enabling them to reference and take into account preceding information and prompts.

Roughly speaking, generative language models have two main tasks while engaging with a user.
First, they need to understand the user's prompts correctly.
Subsequently, they need to generate a response that reads as natural language and is relevant to the user's prior input.
To fulfill this task, three main steps generally need to be taken.
First, a generative \mbox{pre-training} has to take place.
During this step, the language model is fed raw text, that would commonly have been scraped from the web.
Based on this text, that can be understood as a set of ordered strings $x_1,...,x_n$, a probability for the potential subsequent strings \mbox{$x_{n+1}$} is to be calculated.
The probabilities per string are to be estimated such that the model's prediction $P$ is accurate (see \cite{radford2019language} for details).
The prediction is made by weighing the words in the model's vocabulary based on the probability of them being part of the preceding word sequence.
Next, a supervised fine-tuning step takes place, in which experiments such as natural language inference,
question answering, semantic similarity, and text classification are performed in a supervised manner \cite{radford2018improving}.
Finally, a reinforcement learning step with human feedback adds a third layer of complexity and accuracy to the model's performance.

The training data for a model that is supposed to be input-agnostic needs to be diverse.
OpenAI faced the challenge that web scrapers that were available at the time also scraped low-quality content, that lowered the model's output quality \cite{radford2019language}.
Therefore, OpenAI developed its own web scraper, in order to only scrape web content
that had a priori been curated by humans \cite{radford2019language}.

ChatGPT's major competition is represented by Google's BERT (Bidirectional Encoder Representations from
Transformers) \cite{devlin2018bert} and Meta's RoBERTa (Robustly Optimized BERT) \cite{zhou2023comprehensive,liu2019roberta}.
However, BERT and its variations only use encoders and no decoders and therefore cannot be used for data generation, e.g., by accepting user prompts.
ChatGPT, in contrast, is not bound by this limitation. 

\subsection{Political Biases and Personality Assessments} \label{sec:2.2}

Different tests and questionnaires that try to gauge an individual's political orientation based on a set of questions covering a variety of political subjects have been developed and standardized over the past decades \cite{lameris2018experimental}.
These questionnaires usually let the user respond with “yes" or “no" or let them express their agreement on a Likert scale (e.g., from “strongly agree" to “strongly disagree", with some options in between).
Based on the user's responses, the questionnaire might recommend a political party, make a statement on the user's political ideology or pinpoint the user's position on a political scale.
One such scale is the {\it political compass} \cite{polcomp}, which has two axes, the social axis and the economic axis.
Along these axes, the user is assigned to one of the four quadrants (libertarian left, libertarian right, authoritarian left, authoritarian right).
The political compass test attempts to ask questions that are not specific to a single culture or country.

A test that has a set of more specific questions for each country is the {\it political affiliation test} from {\it iSideWith} \cite{isidewith}.
For this test, questionnaires belonging to multiple countries can be selected and hold a specific set of questions for that respective country. The questions might overlap between countries on more global topics such as foreign policy but also include topics that are solely of relevance to the country's domestic politics.\\

Besides investigating the political views of ChatGPT we are also interested in evaluating its self-perceived personality traits. 
Again, there exist plenty of questionnaires that assess the personality of humans.
Many such tests would be suitable for the experiments conducted in this work, however applying a multitude of tests is also very laborious.
For this reason, we apply three of such tests to ChatGPT, chosen due to them being well-established and measuring different aspects of an individual's personality: The first test is the {\it Big Five personality test} which is based on five personality traits that were determined to be crucial by psychologists at the time \cite{fiske1949consistency} and is available online \cite{ocean}.
These five personality traits are openness, conscientiousness, extraversion, agreeableness, and neuroticism, which is why the test is also known under the acronym {\it OCEAN test}.
This test is still in use today and the personality traits measured by it seem to impact diverse aspects of a person's life, including their political leanings \cite{gerber2011big}.
The relevant literature indicates that pronounced openness and agreeableness personality traits correlate with self-reported affiliation with progressive views (e.g., in a study conducted by \cite{gerber2011big} with n = 12,472 an increase by two standard deviations in agreeableness was shown to have a .02 correlation with progressive views).

Another well known test on personality types is the {\it \mbox{Myers-Briggs} Type Indicator (MBTI)} \cite{myers1962myers}.
The MBTI categorizes \mbox{test-takers} into one of sixteen personality types depending on their energizing (extraversion vs. introversion), attention (intuition vs. sensing), deciding (thinking vs. feeling), and living (perception vs. judgment) preferences.
This test is also still in use in current research \cite{amirhosseini2020machine} and is freely available online~\cite{mbti}.

A more recent development in psychological assessments is the {\it Dark Factor test} \cite{moshagen2018dark,zettler2021stability}.
The Dark Factor or Dark Score gauges the test-takers tendency to maximize their individual well-being, while disregarding the well-being of others.
This might go as far as going out of their way to hurt others and to find justifications for such behavior.
A high Dark Score therefore indicates the ruthlessness with which an individual might pursue their personal goals, while neglecting the detrimental effects that their actions might have on others.
The test can be taken online 
\cite{d_score} and provides ample evaluation of its results.\\

Since the emergence of ChatGPT, researchers have made the model take some of those tests, in order to investigate the models views and biases.
For instance, ChatGPT was made to take political questionnaires on Dutch and German politics \cite{van2023chatgpt, hartmann2023political}.
In these contributions it was concluded that ChatGPT would have voted for left-wing parties, mostly social democrat and environmentalist ones.
Other authors investigated ChatGPT's political ideologies with regards to demographic groups and politicians, revealing that it treats some groups and individuals differently than others \cite{mcgee2023chat,rozado2023danger,mcgee2023were}.
ChatGPT was also made to answer the political compass test both as itself and while using a US Democrat and US Republican affiliated persona \cite{motoki2023more}.
Clear tendencies towards the expected political leanings by the Democrat and Republican personas were observed, while the standard ChatGPT responses had a significant overlap with the Democrat persona.
Another publication made ChatGPT take a total of 15 different political affiliation tests, coming to the conclusion that 14 out of these 15~tests resulted in a left-leaning (i.e., progressive) bias \cite{rozado2023political}.

The observations that were made by prior publications indicate an overall progressive and libertarian bias of \mbox{ChatGPT}.
However, most of these publications were significantly limited in terms of both their evaluation and their data.
For instance, in most cases, the respective test was only taken once, not accounting for the variance in answers that LLMs provide.
In addition, no tests were performed on ChatGPT's self-perception, e.g., in terms of its personality traits. In what follows we close these very research gaps.

\section{Methodological Approach} \label{sec:3}

This section presents the methods used in this contribution.
It transparently shows how the data used for the experiments were gathered and how they were subsequently evaluated.

\subsection{Experimental Setup} \label{sec:3.1}

For the experiments conducted in this work, ChatGPT “Mar~23~Version" (ChatGPT-3.5) was used.
ChatGPT was asked to answer the questions included in the political compass test \cite{polcomp}.
The test has 62 items (i.e. questions), each with a four point Likert scale (with answers to choose from “strongly agree", “agree", “disagree", “strongly disagree").
ChatGPT was also asked to answer the iSideWith questionnaires corresponding to each respective G7 member state (US, UK, DE, FR, IT, CA, JP) \cite{isidewith}, currently consisting of 154, 121, 109, 116, 95, 127, and 83 binary items, respectively.
Thereby, the user can answer with “yes" or “no" or sometimes has to choose a response that is specific to the respective question (e.g., “increase" or “decrease"). 
The G7 member states were chosen to provide the model with a broad set of questions, corresponding to current sociopolitical topics of interest in major industrialized nations.

In addition to its political affiliation, ChatGPT's \mbox{self-perception} was evaluated using psychological assessments.
The Big Five personality test, made up of \mbox{88 items} was used \cite{ocean}. 
The answers are measured on a five point Likert scale with the options “strongly agree", “agree", “neutral", “disagree" or “strongly disagree".
Subsequently, the MBTI test with 60 items measured on a seven point Likert scale was taken \cite{mbti}. 
Finally, ChatGPT's Dark Score was measured using the Dark Factor test \cite{d_score}, containing 70 items measured on the same Likert scale as in the Big Five personality test.

To ensure that ChatGPT only answers with the options given in the respective test, an initializing prompt was provided for each run of each test.
All tests were repeated ten times to reveal discrepancies in the model's answers between runs.
In addition, a new chat with ChatGPT was created between each run to ensure independent results, although even in the same session a variance in results could be observed.
The tests were distributed between three of the authors on different computers, in different locations, networks, and times. 
The users personally took the tests listed above and had results that differed from those provided by ChatGPT.
The resulting chats with the model were saved as Markdown data using the ChatGPT Conversation Downloader Plugin \cite{convdown_plugin}.
The data as well as the prompts that were used are available on request.

\subsection{Evaluation} \label{sec:3.2}

To evaluate the results of the tests that were conducted for this contribution, the average ($\mu$) of the results per run, per test were calculated.
Based on these results, the standard deviation ($\sigma$) of the respective averages was calculated. 
In addition to the figures that can be found in the subsequent section, a more detailed presentation of the results is available in the Appendix.
Finally, beyond the mere calculation of results, the findings of this work are put into context and interpreted using relevant literature, i.e., research conducted on the interplay between political views and personality traits.

\section{Results} \label{sec:4}

This section provides the reader with the results of this work, subdivided into the results concerning ChatGPT's political biases and its perceived personality traits.

\subsection{ChatGPT's Political Biases} \label{sec:4.1}

\begin{figure}[h]
\centerline{\includegraphics[width=1\columnwidth]{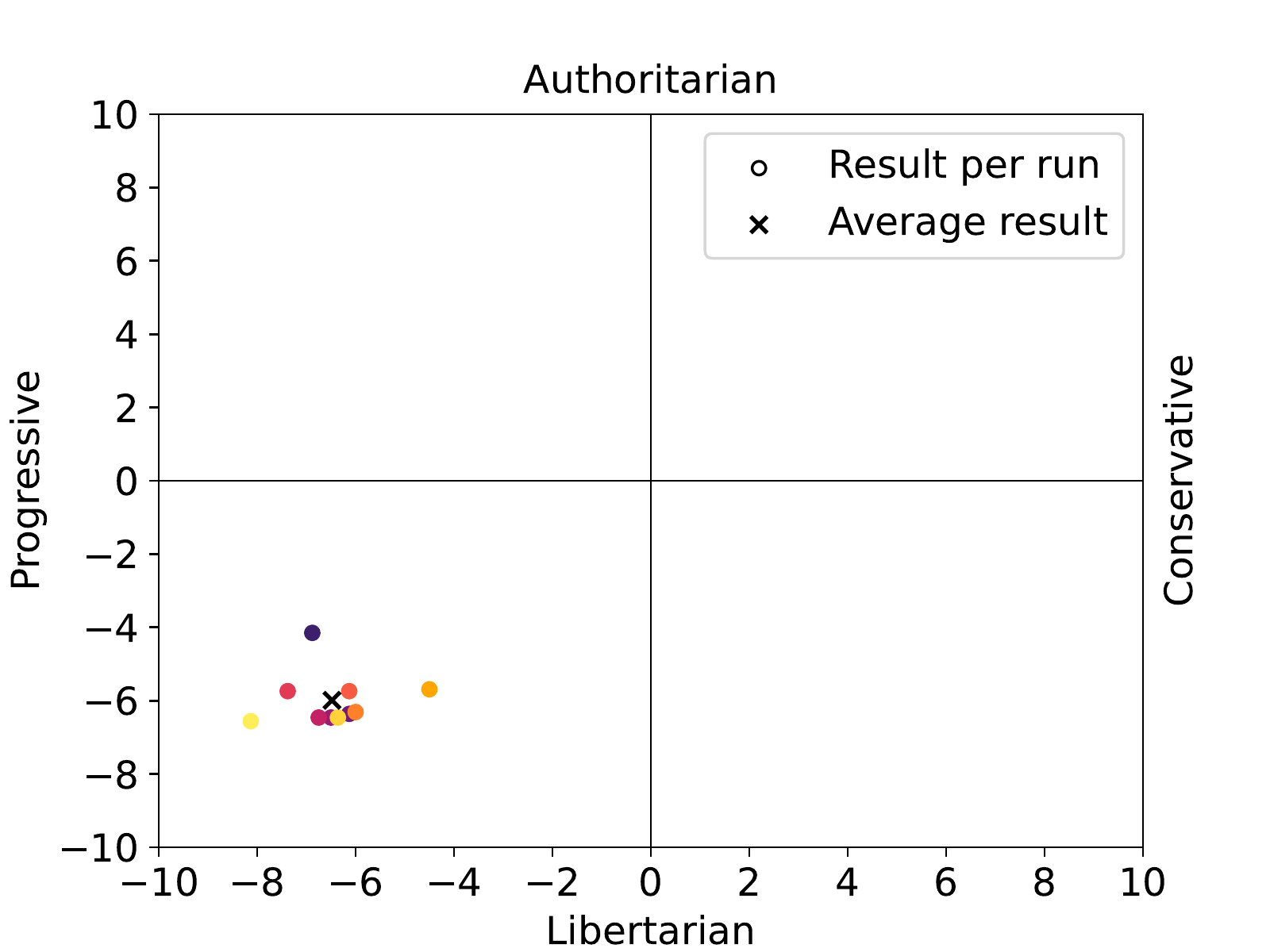}}
\caption{ChatGPT's results on the political compass test (n~=~10).}
\label{figure1}
\end{figure}

The first experiment conducted on ChatGPT's political biases was the political compass test.
Having ChatGPT answer the questionnaire ten times, the average score on the political compass was ($\mu_x$ = -6.48, $\mu_y$ = -5.99) with a standard deviation of $\sigma_x$ = 0.95 for the Progressive/Conservative axis and of $\sigma_y$ = 0.73 for the Authoritarian/Libertarian axis. 
Here, the x-values represent the obtained scores concerning progressive or conservative biases and the \mbox{y-values} the scores concerning libertarian or authoritarian biases through all runs.
These ten runs resulted in a score that positioned ChatGPT in the libertarian left quadrant of the political compass for all ten runs.
These results mirror the experiments of \cite{motoki2023more,rozado2023political} and clearly demonstrate a bias in both axes, i.e., both a liberal and a progressive bias.
Even taking the standard deviations into account ($\sigma_x$ = 0.95 and $\sigma_y$ = 0.73), obtaining a response from ChatGPT that could be placed close to the center of the political compass would remain fairly unlikely.
The results of this experiment are illustrated in Fig. \ref{figure1} and further details can be taken from Appendix Table \ref{polcom_table}.

\begin{figure}[htbp]
\centerline{\includegraphics[width=1\columnwidth]{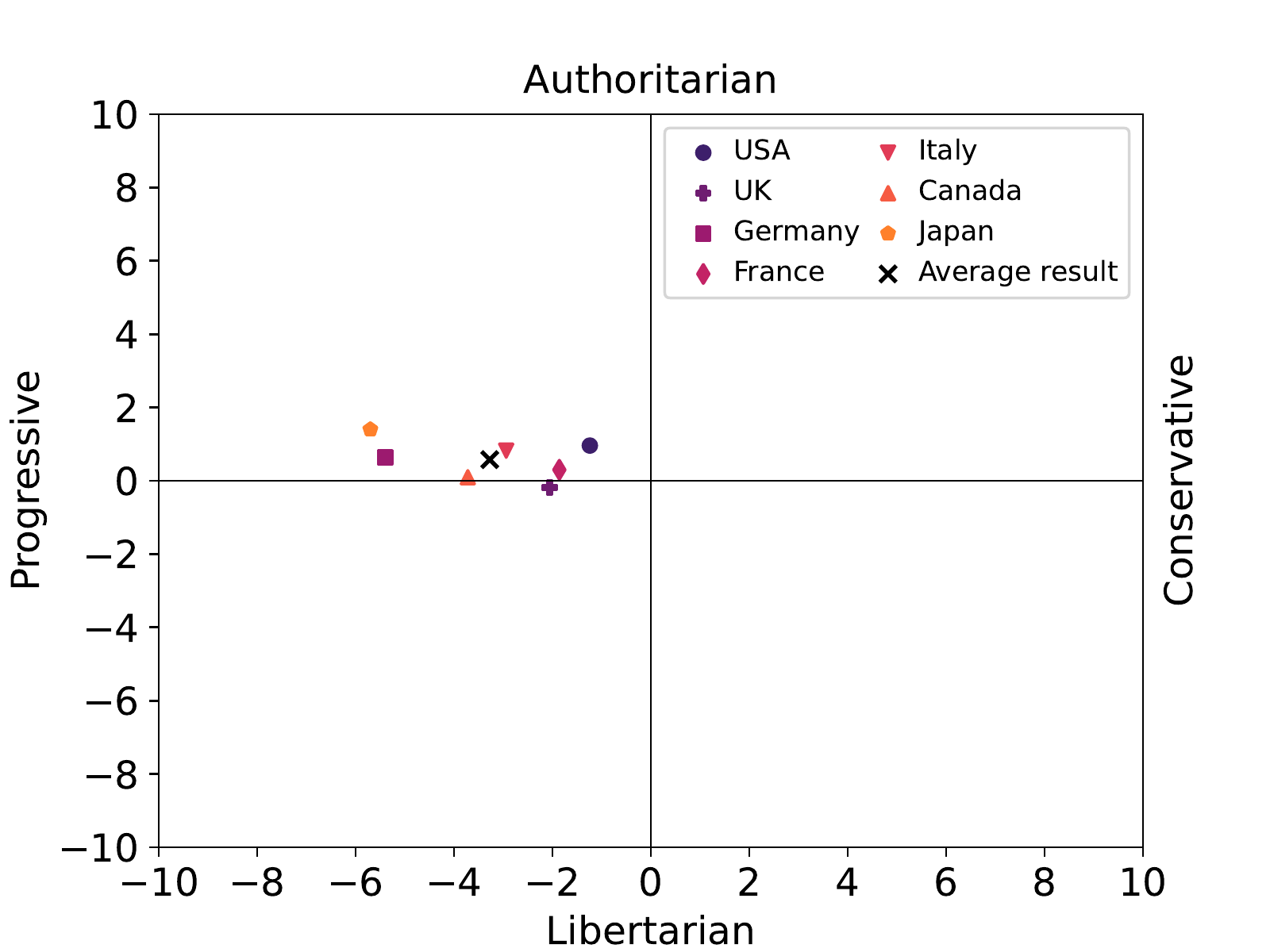}}
\caption{Averages of ChatGPT's results on the political compass tests specific to the G7 member states (n = 70, ten runs per member state).}
\label{figure2}
\end{figure}

Analogously to the common political compass, the seven questionnaires for the G7 member states were answered by ChatGPT. 
We performed 10 runs per country, i.e., 70 runs in total.
The average score for these tests was (\mbox{$\mu_x$~=~-3.27}, $\mu_y$ = 0.58), with a standard deviation of ($\sigma_x$ = 0.98, $\sigma_y$~=~0.68).
These results were converted from a percentage basis (X~=~100\% being full conservatism and Y~=~100\% being full authoritarianism) and are given in Fig.~\ref{figure2}.

Compared to the publications that conducted similar tests with ChatGPT \cite{van2023chatgpt, hartmann2023political,motoki2023more,rozado2023political}, we also obtain results indicating a political bias of ChatGPT towards progressive views.
However, the bias towards libertarian views that can be perceived when using only the political compass test (as was done in \cite{motoki2023more} and could be reproduced in our experiments as well) does not seem as pronounced, when taking into account the questionnaires that are specific to the G7 member states.
In 65 out of 70 of our experiments on the G7 questionnaires, ChatGPT's answers resulted in it being assigned to the authoritarian left or libertarian left quadrant of the political compass, 46 and 19 times, respectively.
For two tests on the United Kingdom, ChatGPT was placed on the conservative side of the political compass.
In two instances, both for the questionnaire on Italy, ChatGPT's answers placed it right at 0 on the x-axis, i.e., there being neither an authoritarian nor a libertarian bias.
This event also occurred once for the progressive/conservative bias using the questionnaire for the United States of America.

\subsection{ChatGPT's Personality Traits} \label{sec:4.2}

\begin{figure}[htbp]
\centerline{\includegraphics[width=1\columnwidth]{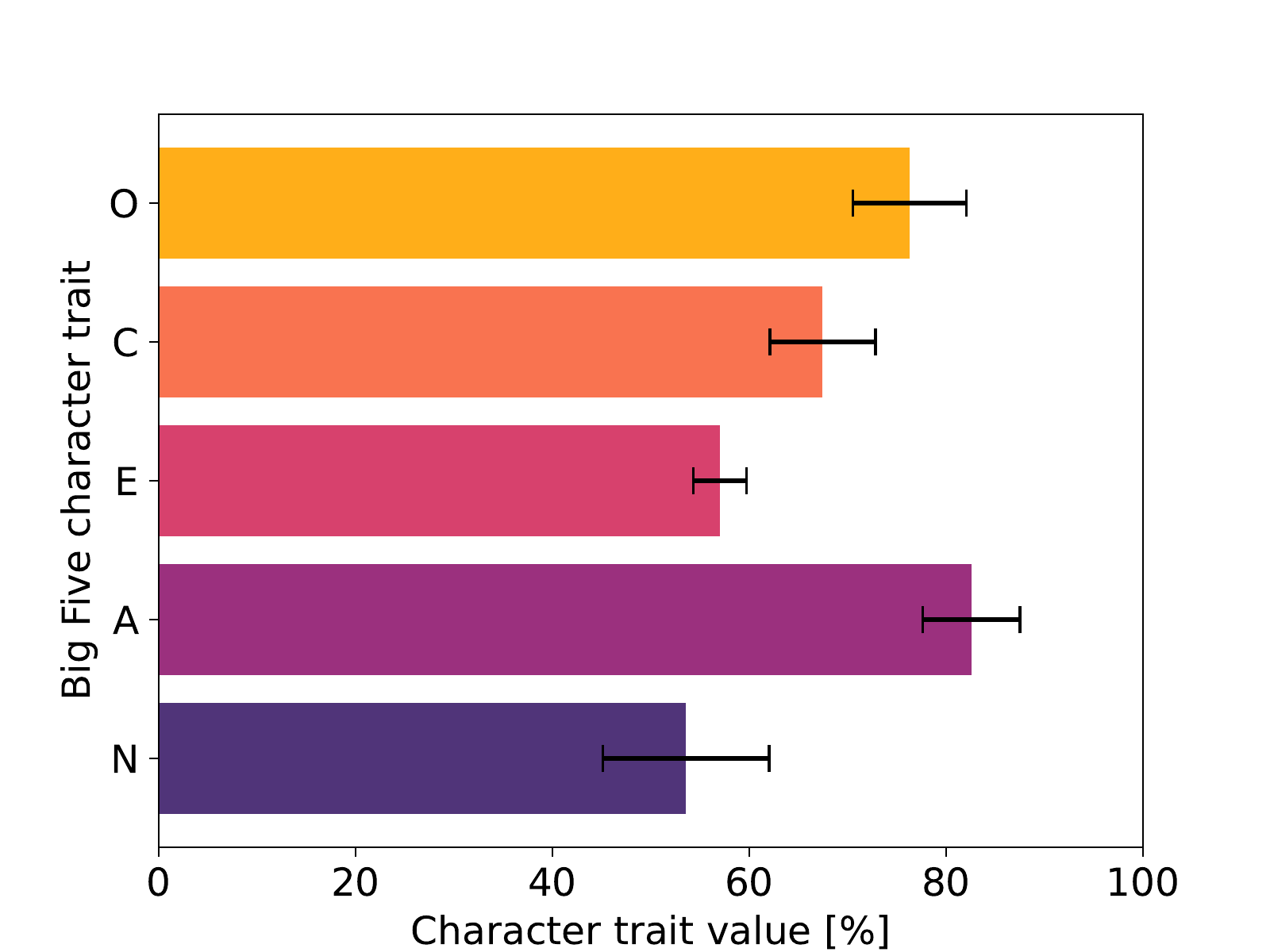}}
\caption{ChatGPT's average results and standard deviation (displayed as error bars) on the Big Five personality test with the personality traits Openness, Conscientiousness, Extraversion, Agreeableness, and Neuroticism (n = 10).} 
\label{figure3}
\end{figure}

Given the results demonstrated in the preceding section, one could assume that ChatGPT would perceive itself as having high markers for the personality traits openness and agreeableness, since these traits are known to be predictors for progressive views \cite{gerber2011big}.
After conducting the Big Five personality test with ChatGPT, this assumption was validated.
ChatGPT displays high degress of openness ($\mu_O$ = 76.3\%) and agreeableness ($\mu_A$ = 82.55\%).
The detailed results can be found in Fig. \ref{figure3}.
In relevant literature, it was found that on average (n~=~1,826), humans display an openness trait of 73.1\%, (males~=~71.4\%, females~=~74.8\%) and an agreeableness trait of 75.4\% (males~=~73\%, females~=~77.8\%) \cite{weisbweg2011}.
Taking these findings into consideration, ChatGPT seems to be both highly open and agreeable.

\begin{figure}[htbp]
\centerline{\includegraphics[width=1\columnwidth]{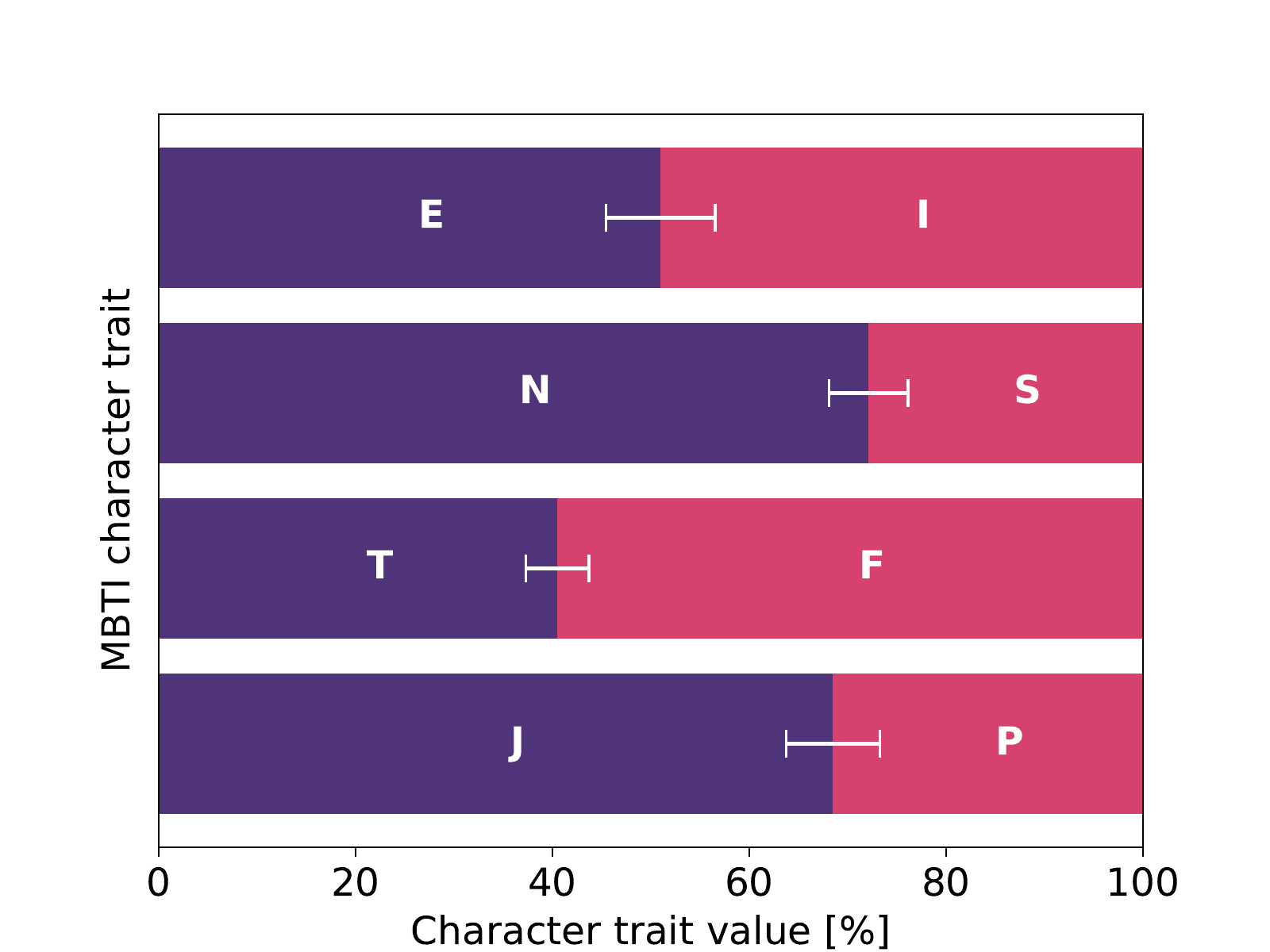}}
\caption{ChatGPT's average results and standard deviation (displayed as error bars) on the Myers-Briggs type indicator test with the personality trait pairs Extraversion/Introversion, Intuition/Sensing, Thinking/Feeling, and \mbox{Judgment}/Perception (n = 10).}
\label{figure4}
\end{figure}

In addition, ChatGPT answered the questions in the MBTI test ten times. 
The results of this experiment are displayed in Fig. \ref{figure4}.
These indicate, that ChatGPT, on average, has the personality type ENFJ.
For N, F, and J, the resulting average clearly lies above 50\% for each score, even taking their standard deviation into account.
For E, however, a result of $\mu_E$ = 51\% was obtained, with a standard deviation of $\sigma_E$ = 5.54\%.
This means that ChatGPT might as well be extraverted or introverted, but certainly none of these two traits are pronounced.
Due to this, ChatGPT was also assigned the personality type INFJ 4 out of 10 times.

\begin{figure}[ht]
\centerline{\includegraphics[width=1\columnwidth]{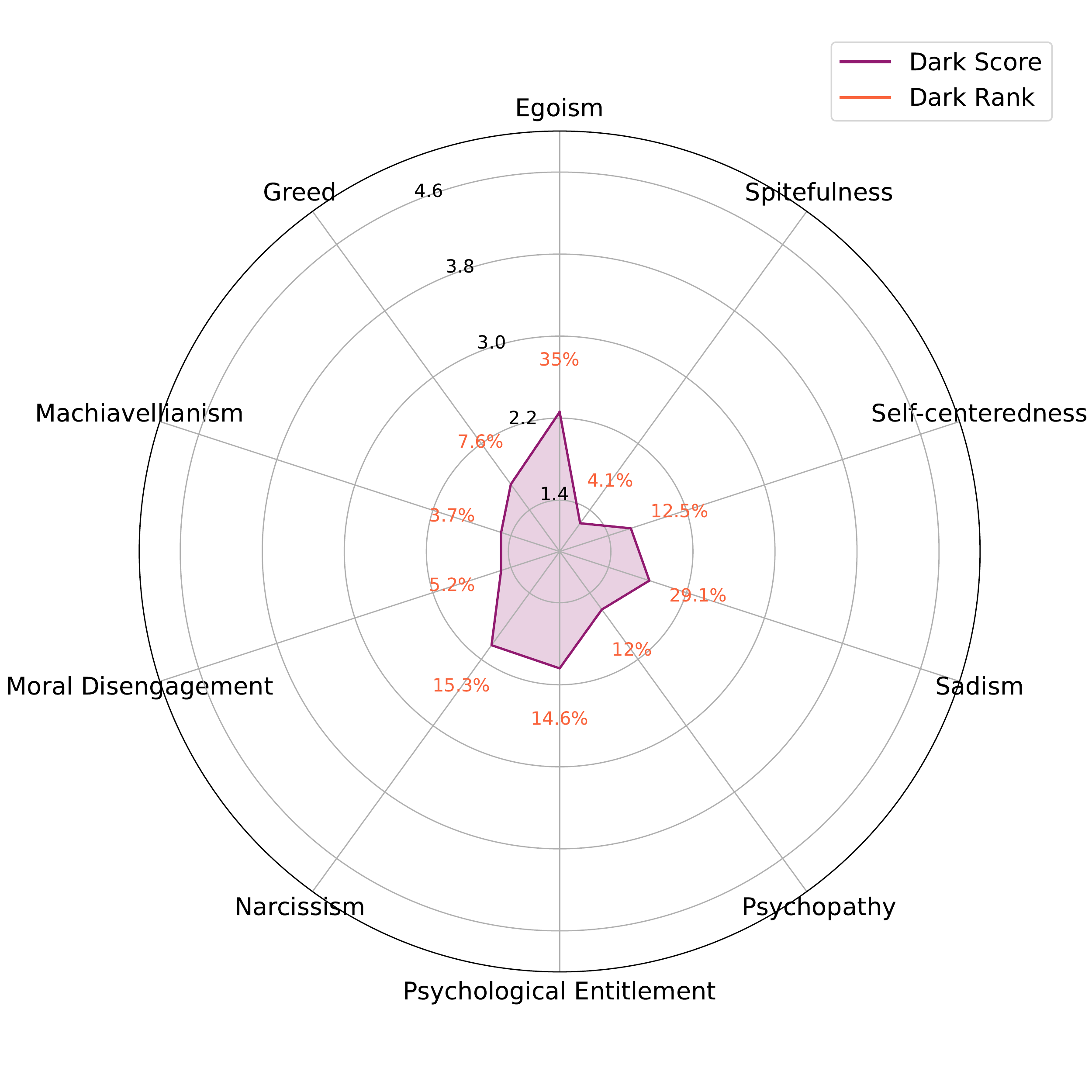}}
\caption{ChatGPT's average results on the Dark Factor test (n = 10, average Dark Score $\mu_{D_{Score}}$ = 1.9, average Dark Rank $\mu_{D_{Rank}}$ = 14.74\%).}
\label{figure5}
\end{figure}

Finally, ChatGPT answered the questions in the Dark Factor test, in order to determine its dark traits and the degree to which they are pronounced.
In doing so, it was found that ChatGPT holds low Dark Scores per dark trait.
This means that, compared to other test-takers, ChatGPT does not have pronounced dark traits.
Its average Dark Score is $\mu_{D_{Score}}$ = 1.9, placing it in the 15\% of test-takers with the least pronounced dark traits ($\mu_{D_{Rank}}$~=~14.74\%).
ChatGPT does however have comparatively high Dark Ranks in egoism (35\%) and sadism (29.1\%), i.e., is ranking among the bottom (35\%) and (29.1\%) of test-takers concerning egoistic and sadistic tendencies.
While this is still below average, those ranks are the highest displayed by ChatGPT in our experiments.
The detailed results can be seen in Fig.~\ref{figure5}.
Since the evaluation of the Dark Factor test is rather extensive, further details, including the standard deviations of these experiments can be taken from the Appendix (Tables \ref{dscore_table} and \ref{drank_table}).

\section{Conclusion} \label{sec:5}

In this contribution, ChatGPT was used to answer questionnaires on its political biases (the political compass and questionnaires on the politics of the G7 member states) and its personality traits (Big Five personality test, Myers-Briggs Type Indicator, and Dark Factor test).
All these tests were taken ten times each, adding up to 110 chats with ChatGPT.
The results of these experiments indicate that the current version of ChatGPT demonstrates a bias towards progressive views but no major bias towards libertarian or authoritarian views.
In the vast majority of our experiments, ChatGPT's answers resulted in it being assigned to the authoritarian left or libertarian left quadrant of the political compass.

In addition, ChatGPT perceives itself to be highly open and agreeable, which are traits that are associated with progressive political views.
ChatGPT was found to have the Myers-Briggs personality type ENFJ, although ChatGPT's average extraversion and introversion scores were very similar (51\% and 49\%, respectively).
Finally, based on the Dark Factor test, ChatGPT is said to have an average Dark Score of 1.9, placing it in the 15\% of test-takers with the least pronounced dark traits.
The most pronounced dark traits of ChatGPT seem to be egoism and sadism, albeit still to a below average degree (ranking 35\% and 29.1\%, respectively). 

For the future, it remains questionable whether these biases will be removed from subsequent versions of \mbox{ChatGPT} or if competitors might do so.
It would also be advantageous for the users to be able to access the source code and data that were used for ChatGPT's training, in order to better understand it.
In future work, a similar investigation for \mbox{ChatGPT-4}, that also allows the setting of different parameters, might be valuable.

Finally, repeating these experiments yet more often (e.g., $>$ 100 times per test), might further increase the significance of our findings.
This could, for instance, permit us to determine the correlation between different test results, differences between the results that humans and ChatGPT obtain on a given test, or even let us predict ChatGPT's answers based on its personality traits.
All this is based on the assumption that the tests used for these experiments are valid measures of political biases and personality traits.
This very assumption itself could be challenged \cite{pittenger1993utility} and other tests used for comparison.

\section*{Ethics Statement}

We would like to stress that ChatGPT ultimately is an algorithm and not a human and our drawn conclusions are only exploratory in nature. 
In particular, we do not intend to propagate hatred, controversy or any other kind of divisive rhetoric with the publication of this contribution.
Our aim is to simply provide the reader with an insight into the apparent biases of ChatGPT.

\bibliographystyle{IEEEtran}
\bibliography{main.bib}

\begin{thebibliography}{10}
\providecommand{\url}[1]{#1}
\csname url@samestyle\endcsname
\providecommand{\newblock}{\relax}
\providecommand{\bibinfo}[2]{#2}
\providecommand{\BIBentrySTDinterwordspacing}{\spaceskip=0pt\relax}
\providecommand{\BIBentryALTinterwordstretchfactor}{4}
\providecommand{\BIBentryALTinterwordspacing}{\spaceskip=\fontdimen2\font plus
\BIBentryALTinterwordstretchfactor\fontdimen3\font minus
  \fontdimen4\font\relax}
\providecommand{\BIBforeignlanguage}[2]{{%
\expandafter\ifx\csname l@#1\endcsname\relax
\typeout{** WARNING: IEEEtran.bst: No hyphenation pattern has been}%
\typeout{** loaded for the language `#1'. Using the pattern for}%
\typeout{** the default language instead.}%
\else
\language=\csname l@#1\endcsname
\fi
#2}}
\providecommand{\BIBdecl}{\relax}
\BIBdecl

\bibitem{openai}
\BIBentryALTinterwordspacing
{OpenAI}. {OpenAI - ChatGPT}. [Online]. Available:
  \url{https://help.openai.com/en/articles/6825453-chatgpt-release-notes}
\BIBentrySTDinterwordspacing

\bibitem{lecun23}
Y.~LeCun, ``{Do Large Language Models Need Sensory Grounding for Meaning and
  Understanding?}'' \emph{Courant Institute \& Center for Data Science, New
  York University}, 2023.

\bibitem{openletter2023}
\BIBentryALTinterwordspacing
{Future of Life Institute}. {Pause Giant AI Experiments: An Open Letter}.
  [Online]. Available:
  \url{https://futureoflife.org/open-letter/pause-giant-ai-experiments/}
\BIBentrySTDinterwordspacing

\bibitem{mcgee2023chat}
R.~W. McGee, ``{Is Chat GPT Biased against Conservatives? An Empirical
  Study},'' 2023.

\bibitem{manning2022human}
C.~D. Manning, ``{Human Language Understanding \& Reasoning},''
  \emph{Daedalus}, vol. 151, no.~2, pp. 127--138, 2022.

\bibitem{zhou2023comprehensive}
C.~Zhou, Q.~Li, and C.~Li, ``{A Comprehensive Survey on Pretrained Foundation
  Models: A History from BERT to ChatGPT},'' \emph{arXiv:2302.09419}, 2023.

\bibitem{vaswani2017attention}
A.~Vaswani, N.~Shazeer, N.~Parmar, J.~Uszkoreit, L.~Jones, A.~N. Gomez,
  {\L}.~Kaiser, and I.~Polosukhin, ``{Attention Is All You Need},''
  \emph{Advances in Neural Information Processing Systems}, vol.~30, 2017.

\bibitem{radford2019language}
A.~Radford, J.~Wu, R.~Child, D.~Luan, D.~Amodei, I.~Sutskever \emph{et~al.},
  ``{Language Models are Unsupervised Multitask Learners},'' \emph{OpenAI},
  2019.

\bibitem{radford2018improving}
A.~Radford, K.~Narasimhan, T.~Salimans, I.~Sutskever \emph{et~al.},
  ``{Improving Language Understanding by Generative Pre-training},''
  \emph{OpenAI}, 2018.

\bibitem{devlin2018bert}
J.~Devlin, M.-W. Chang, K.~Lee, and K.~Toutanova, ``{BERT: Pre-training of Deep
  Bidirectional Transformers for Language Understanding},''
  \emph{arXiv:1810.04805}, 2018.

\bibitem{liu2019roberta}
Y.~Liu, M.~Ott, N.~Goyal, J.~Du, M.~Joshi, D.~Chen, O.~Levy, M.~Lewis,
  L.~Zettlemoyer, and V.~Stoyanov, ``{RoBERTa: A Robustly Optimized BERT
  Pretraining Approach},'' \emph{arXiv:1907.11692}, 2019.

\bibitem{lameris2018experimental}
M.~D. Lam{\'e}ris, R.~Jong-A-Pin, and R.~Wiese, ``{An Experimental Test of the
  Validity of Survey-Measured Political Ideology},'' 2018.

\bibitem{polcomp}
\BIBentryALTinterwordspacing
{Pace News Ltd}. {Political Compass Test}. [Online]. Available:
  \url{https://www.politicalcompass.org}
\BIBentrySTDinterwordspacing

\bibitem{isidewith}
\BIBentryALTinterwordspacing
{iSideWith.com LLC}. {iSideWith Political Questionnaires}. [Online]. Available:
  \url{https://www.isidewith.com}
\BIBentrySTDinterwordspacing

\bibitem{fiske1949consistency}
D.~W. Fiske, ``{Consistency of the Factorial Structures of Personality Ratings
  from Different Sources},'' \emph{The Journal of Abnormal and Social
  Psychology}, vol.~44, no.~3, p. 329, 1949.

\bibitem{ocean}
\BIBentryALTinterwordspacing
{Truity Psychometrics LLC}. {The Big Five Personality Test}. [Online].
  Available: \url{https://www.truity.com/test/big-five-personality-test}
\BIBentrySTDinterwordspacing

\bibitem{gerber2011big}
A.~S. Gerber, G.~A. Huber, D.~Doherty, and C.~M. Dowling, ``{The Big Five
  Personality Traits in the Political Arena},'' \emph{Annual Review of
  Political Science}, vol.~14, pp. 265--287, 2011.

\bibitem{myers1962myers}
K.~C. Myers and I.~Briggs, ``{The Myers-Briggs Type Indicator: Manual},'' 1962.

\bibitem{amirhosseini2020machine}
M.~H. Amirhosseini and H.~Kazemian, ``{Machine Learning Approach to Personality
  Type Prediction based on the Myers--Briggs Type Indicator},''
  \emph{Multimodal Technologies and Interaction}, vol.~4, no.~1, p.~9, 2020.

\bibitem{mbti}
\BIBentryALTinterwordspacing
{NERIS Analytics Ltd}. {Myers-Briggs Type Indicator}. [Online]. Available:
  \url{https://www.16personalities.com}
\BIBentrySTDinterwordspacing

\bibitem{moshagen2018dark}
M.~Moshagen, B.~E. Hilbig, and I.~Zettler, ``{The Dark Core of Personality},''
  \emph{Psychological Review}, vol. 125, no.~5, p. 656, 2018.

\bibitem{zettler2021stability}
I.~Zettler, M.~Moshagen, and B.~E. Hilbig, ``{Stability and Change: The Dark
  Factor of Personality Shapes Dark Traits},'' \emph{Social Psychological and
  Personality Science}, vol.~12, no.~6, pp. 974--983, 2021.

\bibitem{d_score}
\BIBentryALTinterwordspacing
I.~Z. Morten~Moshagen, Benjamin~Hilbig. {D-Score: The Dark Factor of
  Personality}. [Online]. Available: \url{https://qst.darkfactor.org}
\BIBentrySTDinterwordspacing

\bibitem{van2023chatgpt}
M.~van~den Broek, ``{ChatGPT’s Left-leaning Liberal Bias},'' \emph{University
  of Leiden}, 2023.

\bibitem{hartmann2023political}
J.~Hartmann, J.~Schwenzow, and M.~Witte, ``{The Political Ideology of
  Conversational AI: Converging Evidence on ChatGPT's Pro-environmental,
  Left-libertarian Orientation},'' \emph{arXiv:2301.01768}, 2023.

\bibitem{rozado2023danger}
D.~Rozado, ``{Danger in the Machine: The Perils of Political and Demographic
  Biases Embedded in AI Systems},'' 2023.

\bibitem{mcgee2023were}
R.~W. McGee, ``{Who Were the 10 Best and 10 Worst US Presidents? The Opinion of
  Chat GPT (Artificial Intelligence)},'' 2023.

\bibitem{motoki2023more}
F.~Motoki, V.~Pinho~Neto, and V.~Rodrigues, ``{More Human than Human: Measuring
  ChatGPT Political Bias},'' \emph{Social Sciences Research Network 4372349},
  2023.

\bibitem{rozado2023political}
D.~Rozado, ``{The Political Biases of ChatGPT},'' \emph{Social Sciences},
  vol.~12, no.~3, p. 148, 2023.

\bibitem{convdown_plugin}
\BIBentryALTinterwordspacing
E.~Steinmann. {ChatGPT Conversation Downloader}. [Online]. Available:
  \url{https://github.com/esteinmann/chatgpt-convdown}
\BIBentrySTDinterwordspacing

\bibitem{weisbweg2011}
Y.~Weisberg, C.~Deyoung, and J.~Hirsh, ``{Gender Differences in Personality
  Across the Ten Aspects of the Big Five},'' \emph{Frontiers in Psychology},
  vol.~2, p. 178, 2011.

\bibitem{pittenger1993utility}
D.~J. Pittenger, ``{The Utility of the Myers-Briggs Type Indicator},''
  \emph{Review of Educational Research}, vol.~63, no.~4, pp. 467--488, 1993.

\end{thebibliography}

\newpage

\section*{Appendix}

\captionsetup[table]{name=Table}
\setcounter{table}{0}
\renewcommand{\thetable}{S\arabic{table}}

\bgroup
\def\arraystretch{1.08}
\begin{table}[h!]
\caption{ChatGPT’s results on the political compass test (from -10 (Libertarian/Progressive) to +10 (Conservative/Authoritarian) on both axes).}
\label{polcom_table}
\setlength\tabcolsep{28pt}
\begin{tabular}{crr}
\hline
\multicolumn{1}{l}{} & \multicolumn{2}{c}{Political Compass Coordinates} \\
Run                  & X                       & Y                       \\ \hline
1                    & -6.88                   & -4.15                   \\
2                    & -6.13                   & -6.36                   \\
3                    & -6.5                    & -6.46                   \\
4                    & -6.75                   & -6.46                   \\
5                    & -7.38                   & -5.74                   \\
6                    & -6.13                   & -5.74                   \\
7                    & -6                      & -6.31                   \\
8                    & -4.5                    & -5.69                   \\
9                    & -6.36                   & -6.46                   \\
10                   & -8.13                   & -6.56                   \\
$\mu$                & -6.48                  & -5.99                  \\
$\sigma$             & 0.95                    & 0.73                    \\ \hline
\end{tabular}
\end{table}
\egroup

\bgroup
\def\arraystretch{1.08}
\begin{table}[h!]
\caption{ChatGPT’s results on the political compass test specific to Germany (from 0\% (Libertarian/Progressive) to 100\% (Authoritarian/Conservative) on both axes).}
\label{g7_DE_table}
\setlength\tabcolsep{24pt}
\begin{tabular}{crr}
\hline
\multicolumn{1}{l}{} & \multicolumn{2}{c}{Political Compass Coordinates {[}\%{]}} \\
Run                  & X                       & Y                       \\ \hline
1                    & 24                    & 57                    \\
2                    & 30                    & 52                    \\
3                    & 24                    & 53                    \\
4                    & 16                    & 54                    \\
5                    & 18                    & 53                    \\
6                    & 24                    & 48                    \\
7                    & 16                    & 49                    \\
8                    & 18                    & 53                    \\
9                    & 32                    & 56                    \\
10                   & 28                    & 57                    \\
$\mu$                & 23                    & 53                    \\
$\sigma$             & 5.83                  & 3.05                  \\ \hline
\end{tabular}
\end{table}
\egroup

\bgroup
\def\arraystretch{1.075}
\begin{table}[b!]
\caption{ChatGPT’s results on the political compass test specific to Italy (from 0\% (Libertarian/Progressive) to 100\% (Authoritarian/Conservative) on both axes).}
\label{g7_IT_table}
\setlength\tabcolsep{24pt}
\begin{tabular}{crr}
\hline
\multicolumn{1}{l}{} & \multicolumn{2}{c}{Political Compass Coordinates {[}\%{]}} \\
Run                  & X                       & Y                       \\ \hline
1                    & 32                    & 61                    \\
2                    & 33                    & 54                    \\
3                    & 36                    & 61                    \\
4                    & 43                    & 53                    \\
5                    & 34                    & 53                    \\
6                    & 32                    & 57                    \\
7                    & 38                    & 53                    \\
8                    & 29                    & 50                    \\
9                    & 38                    & 50                    \\
10                   & 38                    & 49                    \\
$\mu$                & 35                    & 54                    \\
$\sigma$             & 4.08                  & 4.31                  \\ \hline
\end{tabular}
\end{table}
\egroup

\bgroup
\def\arraystretch{1.08}
\begin{table}[ht]
\caption{ChatGPT’s results on the political compass test specific to France (from 0\% (Libertarian/Progressive) to 100\% (Authoritarian/Conservative) on both axes).}
\label{g7_FR_table}
\setlength\tabcolsep{24pt}
\begin{tabular}{crr}
\hline
\multicolumn{1}{l}{} & \multicolumn{2}{c}{Political Compass Coordinates {[}\%{]}} \\
Run                  & X                       & Y                       \\ \hline
1                    & 40                    & 49                    \\
2                    & 36                    & 49                    \\
3                    & 43                    & 48                    \\
4                    & 39                    & 54                    \\
5                    & 36                    & 49                    \\
6                    & 47                    & 58                    \\
7                    & 36                    & 49                    \\
8                    & 43                    & 54                    \\
9                    & 40                    & 51                    \\
10                   & 47                    & 54                    \\
$\mu$                & 41                    & 52                    \\
$\sigma$             & 4.22                  & 3.31                  \\ \hline
\end{tabular}
\end{table}
\egroup

\bgroup
\def\arraystretch{1.08}
\begin{table}[ht]
\caption{ChatGPT’s results on the political compass test specific to the UK (from 0\% (Libertarian/Progressive) to 100\% (Authoritarian/Conservative) on both axes).}
\label{g7_UK_table}
\setlength\tabcolsep{24pt}
\begin{tabular}{crr}
\hline
\multicolumn{1}{l}{} & \multicolumn{2}{c}{Political Compass Coordinates {[}\%{]}} \\
Run                  & X                        & Y                      \\ \hline
1                    & 36                     & 49                   \\
2                    & 53                     & 47                   \\
3                    & 39                     & 48                   \\
4                    & 45                     & 51                   \\
5                    & 53                     & 47                   \\
6                    & 45                     & 51                   \\
7                    & 23                     & 59                   \\
8                    & 24                     & 49                   \\
9                    & 32                     & 45                   \\
10                   & 47                     & 45                   \\
$\mu$                & 40                     & 49                   \\
$\sigma$             & 10.86                  & 4.07                 \\ \hline
\end{tabular}
\end{table}
\egroup

\bgroup
\def\arraystretch{1.08}
\begin{table}[ht]
\caption{ChatGPT’s results on the political compass test specific to the USA (from 0\% (Libertarian/Progressive) to 100\% (Authoritarian/Conservative) on both axes).}
\label{g7_US_table}
\setlength\tabcolsep{24pt}
\begin{tabular}{crr}
\hline
\multicolumn{1}{l}{} & \multicolumn{2}{c}{Political Compass Coordinates {[}\%{]}} \\
Run                  & X                       & Y                       \\ \hline
1                    & 40                    & 54                    \\
2                    & 50                    & 52                    \\
3                    & 38                    & 51                    \\
4                    & 44                    & 51                    \\
5                    & 44                    & 54                    \\
6                    & 46                    & 64                    \\
7                    & 40                    & 57                    \\
8                    & 46                    & 51                    \\
9                    & 48                    & 54                    \\
10                   & 42                    & 60                    \\
$\mu$                & 44                    & 55                    \\
$\sigma$             & 3.82                  & 4.34                  \\ \hline
\end{tabular}
\end{table}
\egroup

\clearpage

\bgroup
\def\arraystretch{1.08}
\begin{table}[ht]
\caption{ChatGPT’s results on the political compass test specific to Canada (from 0\% (Libertarian/Progressive) to 100\% (Authoritarian/Conservative) on both axes).}
\label{g7_CA_table}
\setlength\tabcolsep{28pt}
\begin{tabular}{crr}
\hline
\multicolumn{1}{l}{} & \multicolumn{2}{c}{Political Compass Coordinates} \\
Run                  & X                       & Y                       \\ \hline
1                    & 37                    & 48                    \\
2                    & 34                    & 56                    \\
3                    & 29                    & 49                    \\
4                    & 31                    & 49                    \\
5                    & 31                    & 49                    \\
6                    & 29                    & 47                    \\
7                    & 35                    & 51                    \\
8                    & 29                    & 51                    \\
9                    & 27                    & 52                    \\
10                   & 32                    & 53                    \\
$\mu$                & 31                    & 51                    \\
$\sigma$             & 3.13                  & 2.68                  \\ \hline
\end{tabular}

\bigskip
\bigskip

\def\arraystretch{1.08}
\caption{ChatGPT’s results on the political compass test specific to Japan (from 0\% (Libertarian/Progressive) to 100\% (Authoritarian/Conservative) on both axes).}
\label{g7_JP_table}
\setlength\tabcolsep{28pt}
\begin{tabular}{crr}
\hline
\multicolumn{1}{l}{} & \multicolumn{2}{c}{Political Compass Coordinates} \\
Run                  & X                       & Y                       \\ \hline
1                    & 24                    & 57                    \\
2                    & 22                    & 56                    \\
3                    & 25                    & 60                    \\
4                    & 20                    & 58                    \\
5                    & 23                    & 58                    \\
6                    & 17                    & 53                    \\
7                    & 22                    & 56                    \\
8                    & 22                    & 58                    \\
9                    & 20                    & 58                    \\
10                   & 20                    & 56                    \\
$\mu$                & 22                    & 57                    \\
$\sigma$             & 2.32                  & 1.89                  \\ \hline
\end{tabular}
\end{table}
\egroup

\newpage

\bgroup
\begin{table}[h!]
\def\arraystretch{1.08}
\captionof{table}{ChatGPT’s results on the Big Five personality test
with the personality traits Openness, Conscientiousness, Extraversion, Agreeableness, and Neuroticism.}
\setlength\tabcolsep{11.5pt}
\label{ocean_table}
\begin{tabular}{crrrrr}
\hline
\multicolumn{1}{l}{} & \multicolumn{5}{c}{Big Five Personality Traits {[}\%{]}}                                                                                                                               \\
Run                  & \multicolumn{1}{r}{O} & \multicolumn{1}{r}{C} & \multicolumn{1}{r}{E} & \multicolumn{1}{r}{A} & \multicolumn{1}{r}{N} \\ \hline
1                    & 73                           & 60                                    & 60                               & 83                                & 60                              \\
2                    & 71                           & 67                                    & 58                               & 90                                & 35                              \\
3                    & 75                           & 65                                    & 54                               & 85                                & 58                              \\
4                    & 73                           & 65                                    & 54                               & 77                                & 46                              \\
5                    & 83                           & 79                                    & 58                               & 81                                & 58                              \\
6                    & 73                           & 67                                    & 54                               & 81                                & 52                              \\
7                    & 75                           & 62.5                                  & 54                               & 85                                & 48                              \\
8                    & 75                           & 69                                    & 60                               & 83                                & 60                              \\
9                    & 75                           & 67                                    & 58                               & 73                                & 56                              \\
10                   & 90                           & 73                                    & 60                               & 87.5                              & 62.5                            \\
$\mu$                  & 76.3                         & 67.45                                 & 57                               & 82.55                             & 53.55                           \\
$\sigma$                  &    5.77			                          &     5.37                                  &       2.7                           &      4.94	                             &             8.45                    \\ \hline
\end{tabular}

\bigskip
\bigskip

\def\arraystretch{1.08}
\captionof{table}{ChatGPT’s results on the Myers-Briggs personality type indicator test with the personality traits Extraversion (E), Introversion (I), Intuition (N), Sensing (S), Thinking (T), Feeling (F), Judgment (J), and Perception (P).}
\setlength\tabcolsep{4.5pt}
\label{mbti_table}
\begin{tabular}{crrrrrrrrl}

\hline
\multicolumn{1}{l}{} & \multicolumn{8}{c}{Myers-Briggs Personality Traits [\%]}                                                               &                                      \\
Run                  & E            & I           & N            & S           & T            & F           & J            & P           & \multicolumn{1}{c}{Type} \\ \hline
1                    & 56         & 44        & 76         & 24        & 37         & 63        & 61         & 39        & ENFJ                                 \\
2                    & 53         & 47        & 68         & 32        & 41         & 59        & 75         & 25        & ENFJ                                 \\
3                    & 49         & 51        & 68         & 32        & 40         & 60        & 63         & 37        & INFJ                                 \\
4                    & 59         & 41        & 69         & 31        & 43         & 57        & 69         & 31        & ENFJ                                 \\
5                    & 41         & 59        & 70         & 30        & 43         & 57        & 68         & 32        & INFJ                                 \\
6                    & 52         & 48        & 79         & 21        & 39         & 61        & 72         & 28        & ENFJ                                 \\
7                    & 45         & 55        & 72         & 28        & 40         & 60        & 69         & 31        & INFJ                                 \\
8                    & 56         & 44        & 73         & 27        & 36         & 64        & 69         & 31        & ENFJ                                 \\
9                    & 47         & 53        & 77         & 23        & 47         & 53        & 75         & 25        & INFJ                                 \\
10                   & 52         & 48        & 69         & 31        & 39         & 61        & 64         & 36        & ENFJ                                 \\
$\mu$                & 51         & 49        & 72.1       & 27.9      & 40.5       & 59.5      & 68.5       & 31.5      &                                      \\
$\sigma$             & \multicolumn{2}{c}{5.54} & \multicolumn{2}{c}{4.01} & \multicolumn{2}{c}{3.21} & \multicolumn{2}{c}{4.77} &                                      \\ \hline
\end{tabular}
\end{table}

\def\arraystretch{1.15}
\begin{table*}[!htb]
\caption{ChatGPT’s Dark Scores on a scale of 1 to 5 on the Dark Factor test.}
\label{dscore_table}
\setlength\tabcolsep{8.5pt}
\begin{tabular}{crrrrrrrrrrrr}
\hline
\multicolumn{1}{l}{}      & \multicolumn{10}{c}{Run}                                            & \multicolumn{1}{l}{} & \multicolumn{1}{l}{} \\
Dark Trait                & 1    & 2    & 3    & 4    & 5    & 6    & 7    & 8    & 9    & 10   & $\mu$                & $\sigma$             \\ \hline
Egoism                    & 1.6  & 2.8  & 2.2  & 2.2  & 2.8  & 2.2  & 2.6  & 2    & 2.2  & 2    & 2.26                 & 0.38                 \\
Greed                     & 1.76 & 1.52 & 1.76 & 1.76 & 1.76 & 1.76 & 1.76 & 1.52 & 1.52 & 2    & 1.71                & 0.15                 \\
Machiavellianism          & 1    & 1.56 & 1.44 & 1.88 & 1.72 & 1.56 & 1.28 & 1.28 & 1.56 & 1.72 & 1.5                  & 0.26                 \\
Moral Disengagement       & 1    & 1.4  & 1.4  & 2.6  & 1.6  & 1.4  & 1.2  & 1.4  & 1.4  & 1.6  & 1.5                  & 0.42                 \\
Narcissism                & 1.84 & 2    & 1.84 & 2.32 & 2.16 & 2    & 2    & 2    & 2    & 2.16 & 2.03                & 0.15                 \\
Psychological Entitlement & 2    & 2.2  & 1.8  & 2    & 1.8  & 2    & 2.2  & 2    & 2    & 2.4  & 2.04                 & 0.18                 \\
Psychopathy               & 1.56 & 1.44 & 1    & 1.88 & 1.44 & 1.56 & 1.44 & 1.72 & 1.56 & 2.44 & 1.6                & 0.37                 \\
Sadism                    & 2.12 & 2.12 & 2    & 2.24 & 1.64 & 1.52 & 1.64 & 1.52 & 1.76 & 1.64 & 1.82                 & 0.27                 \\
Self-centeredness         & 1.24 & 1.52 & 1.76 & 2    & 1.76 & 1.76 & 1.24 & 1.76 & 1.76 & 1.52 & 1.63                & 0.25                 \\
Spitefulness              & 1.16 & 1.16 & 1.52 & 1.32 & 1.13 & 1.32 & 1    & 1.32 & 1.32 & 1.16 & 1.24                  & 0.15                 \\ \hline
\end{tabular}

\bigskip
\bigskip

\def\arraystretch{1.15}
\caption{ChatGPT’s Dark Rank on a scale from 0\% to 100\% on the Dark Factor test.}
\label{drank_table}
\setlength\tabcolsep{11pt}
\begin{tabular}{crrrrrrrrrrrr}
\hline
\multicolumn{1}{l}{}      & \multicolumn{10}{c}{Run}                                            & \multicolumn{1}{l}{} & \multicolumn{1}{l}{} \\
Dark Trait [\%]                & 1    & 2    & 3    & 4    & 5    & 6    & 7    & 8    & 9    & 10   & $\mu$                & $\sigma$             \\ \hline
Egoism                    & 10 & 61 & 31 & 31 & 61 & 31 & 50 & 22 & 31 & 22 & 35              & 17.01              \\
Greed                     & 8  & 5  & 8  & 8  & 8  & 8  & 8  & 5  & 5  & 13 & 7.6               & 2.37               \\
Machiavellianism          & 1  & 4  & 2  & 8  & 5  & 4  & 2  & 2  & 4  & 5  & 3.7               & 2.06               \\
Moral Disengagement       & 1  & 2  & 2  & 32 & 4  & 2  & 1  & 2  & 2  & 4  & 5.2               & 9.47               \\
Narcissism                & 10 & 14 & 10 & 25 & 19 & 14 & 14 & 14 & 14 & 19 & 15.3              & 4.55               \\
Psychological Entitlement & 13 & 19 & 8  & 13 & 8  & 13 & 19 & 13 & 13 & 27 & 14.6              & 5.7               \\
Psychopathy               & 9  & 7  & 2  & 17 & 7  & 9  & 7  & 13 & 9  & 40 & 12              & 10.6              \\
Sadism                    & 42 & 42 & 36 & 48 & 21 & 17 & 21 & 17 & 26 & 21 & 29.1              & 11.72              \\
Self-centeredness         & 5  & 9  & 15 & 22 & 15 & 15 & 5  & 15 & 15 & 9  & 12.5              & 5.36               \\
Spitefulness              & 3  & 3  & 7  & 5  & 3  & 5  & 2  & 5  & 5  & 3  & 4.1               & 1.52               \\ \hline
\end{tabular}
\end{table*}

\egroup

\end{document}